\documentstyle[epsf,twoside]{article}
\def\bchanged{\bgroup\bf}
\def\echanged{\egroup}

\newcommand{\eps}[3]{\begin{figure}
\begin{center}\leavevmode\epsfbox{#1}\caption[{\em #2}]%
{\label{#3}{\em #2}}\end{center}
\end{figure}}

\newcommand{\emptyeps}[3]{\begin{figure}\caption[{\em #2}]{\label{#3}{\em
#2}}\end{figure}}
\begin{document}
\thispagestyle{empty}
\bibliographystyle{plain}
\title{Two Lane Traffic Simulations \\ using \\ Cellular Automata\\
{}~\\
{\normalsize LANL Report No.\ LA-UR 95-4367}}
\author{%
M.\ Rickert${}^{a,b}$, K.\ Nagel${}^{b}$, M.\ Schreckenberg${}^{c}$,
A.\ Latour${}^{d}$\\
{}~\\
\footnotesize
${}^a$ Center for Parallel Computing, Universit\"at zu K\"oln, 50923
K\"oln, Germany,\\
\footnotesize
mr@zpr.uni-koeln.de\\
\footnotesize
${}^b$ Los Alamos National Laboratory, TSA-DO/SA MS M997, Los Alamos
NM 87545,\\
\footnotesize
USA, rickert@tsasa.lanl.gov and kai@lanl.gov\\
\footnotesize
${}^c$ Theoretische Physik/FB 10, Gerhard-Mercator-Universit\"at
Duisburg,\\
\footnotesize
47048 Duisburg, Germany, schreck@rs1.comphys.uni-duisburg.de\\
\footnotesize
${}^d$ Institut f\"ur Theoretische Physik, Universit\"at zu K\"oln,
50923 K\"oln,\\
\footnotesize
Germany, a2015846@smail.rrz.uni-koeln.de
}
\maketitle
\begin{abstract}
We examine a simple two lane cellular automaton based upon the single
lane CA introduced by Nagel and Schreckenberg. We point out
important parameters defining the shape of the fundamental diagram. Moreover we
investigate the importance of stochastic elements with respect to real
life traffic.
\end{abstract}

\section{Introduction}

In recent times cellular automata based simulations of traffic flow
have gained considerable importance. By extending the range of rules
from nearest neighbours to a range of 5 grid sites and introducing 6
discrete velocities $0\ldots 5$ Nagel and Schreckenberg
\cite{ NaSchr:1} have found a striking resemblance of simulation and
realistic traffic behaviour. For $v_{max}=1$ Schadschneider and
Schreckenberg have found an analytic solution \cite{ SchaSchr:1}. For
higher $v_{max}$, these analytic approaches lead to good
approximations for the average behavior~\cite{ Schreck.etc}.  Further
analytic results can be found in~\cite{ Brankov...Schreck}.
Nagel~\cite{ Nagel.95.flow} pointed out the strong connections between
particle hopping models and fluid-dynamical approaches for traffic
flow.

Much less is known about modelling of multi-lane traffic.  This
statement is not only true for particle hopping models for traffic
flow, but for traffic flow theory in general.
Queueing models are not truly multilane, but emulate multiple lanes by
switching the order of vehicles on one lane whenever a passing would
have occurred in reality \cite{ albrca:1}.
Fluid-dynamical models incorporate multi-lane traffic only by
parametrization~\cite{ mupi:1, muhsla:1}, although sometimes based on
kinetic theory~\cite{ Helbing.kinetic}.
Traditional car-following theory (see~\cite{ Gerlough}) by and large
never dealt with multi-lane traffic.  Modern microscopic traffic
simulation models (e.g.~\cite{ Benz.AS, FRESIM, Cameron.PARAMICS,
McDonald.Brackstone}) obviously handle multi-lane traffic by
necessity.  Cremer and coworkers~\cite{ Cremer.Ludwig, Schuett} even
treat multi-lane traffic in the context of cellular automata models.
Yet, all these papers approach the problem by using heuristic rules of
human behavior, without checking which of these rules exactly cause
which kind of behavior.  In validations then (e.g.~\cite{
McDonald.Brackstone}), it often enough turns out that certain features
of the model are not realistic; and because of the heuristic approach
it is difficult to decide which rules have to be changed or added in
order to correct the problem.

For that reason, a more systematic approach seems justified.  Our
approach here is to search for {\em minimal\/} sets of rules which
reproduce certain macroscopic facts.  The advantage is that relations
between rules and macroscopic behavior can be more easily identified;
and as a welcome side-effect one also obtains higher computational
speed.

We again choose particle hopping models as starting point for this
investigation because their highly discrete nature reduces the number
of free parameters even further.  It is clear that a similar analysis
could be applied to continuous microscopic models, hopefully
benefiting from the results obtained in this and following papers.

Nagatani examined a two lane system with completely deterministic
rules and $v_{max}=1$~\cite{ nagatani:det1, nagatani:det2}, where cars
either move forward or change lanes.  A very unrealistic feature of
this model are states where blocks of several cars oscillate between
lanes without moving forward at all.  He corrected this by introducing
randomness into the lane changing~\cite{ nagatani:pchange}.
Latour has developed the two lane model which served as the basis for
the one discussed here \cite{ latour:1}.
Rickert used a more elaborate rule set for two lane traffic which
reproduced the phenomenon of increased flow with an imposed speed
limit \cite{ rickert:1}.

\section{Single Lane Model}

For the convenience of the reader we would like to outline the single
lane model. The system consists of a one dimensional grid of $L$ sites
with periodic boundary conditions. A site can either be empty, or
occupied by a vehicle of velocity zero to $v_{max}$. The velocity is
equivalent to the number of sites that a vehicle advances in one
update --- provided that there are no obstacles ahead. Vehicles move
only in one direction.  The index $i$ denotes the number of a vehicle,
$x(i)$ its position, $v(i)$ its current velocity, $v_d(i)$ its maximum
speed\footnote{Note that in the original model all vehicles had the
same maximum velocity $v_{max}$. We now allow for different
\underline{d}esired velocities $v_d(i)$ to include an inhomogeneous
fleet}, $pred(i)$ the number of the preceding\footnote{{\em A
precedes B} in this context means that {\em A} is followed by {\em B}}
vehicle, $gap(i):=x(pred(i))-x(i)-1$ the width of the gap to the
predecessor. At the beginning of each time step the rules are applied
to all vehicles simultaneously (parallel update, in contrast to
sequential updates which yield slightly different results).  Then the
vehicles are advanced according to their new velocities.
\begin{itemize}

\item
{\bf IF} $v(i) \ne v_d(i)$ {\bf THEN} $v(i) := v(i)+1\quad$ ({\bf S1})

\item
{\bf IF} $v(i)>gap(i)$ {\bf THEN} $v(i) := gap(i)\quad$ ({\bf S2})

\item
{\bf IF} $v(i)>0$ {\bf AND} $rand<p_d(i)$ {\bf THEN} $v(i) := v(i)-1\quad$
({\bf S3})

\end{itemize}
{\bf S1} represents a linear acceleration until the vehicle has reached its
maximum velocity $v_d$. {\bf S2} ensures that vehicles having predecessors in
their way slow down in order not to run into them. In {\bf S3} a random
generator
is used to decelerate a vehicle with a certain probability modelling
erratic driver behaviour. The free--flow average velocity
is $v_{max}-p_d$ (for $p_d\ne 1$).

\section{A Generic Two Lane Model}

The single lane model is not capable of modelling realistic traffic
mainly for one reason: a realistic fleet is usually composed of
vehicles types having different desired velocities. Introducing such
different vehicle types in the single lane model only results in {\em
platooning} with slow vehicles being followed by faster ones and the
average velocity reduced to the free--flow velocity of the slowest
vehicle~\cite{ Nagatani.bunching, Ben.Naim}.

We introduce a two lane model\footnote{Note that the results presented
here still concentrate on a single desired velocity
only. The effects of different desired velocities will
be investigated in future papers.} consisting of two parallel single lane
models with periodic boundary conditions and four additional rules
defining the exchange of vehicles between the lanes.
The update step is split into two sub-steps:
\begin{enumerate}

\item
Check the exchange of vehicles between the two lanes according to the
new rule set. Vehicles are only moved {\em sideways}. They do not {\em
advance}.  Note
that in reality this sub-step regarded by itself seems unfeasible since
vehicles are usually incapable of purely transversal motion. Only
together with the second sub-step our update rules make physically
sense.

This first sub-step is implemented as strict parallel update with
each vehicle making its decision based upon the configuration at the
beginning of the time step.

\item
Perform independent single lane updates on both lanes according
to the single lane update rules. In this second sub-step the resulting
configuration of the first sub-step is used.

\end{enumerate}

\pagebreak[3]
\noindent

The most important parameters of the two lane model are as follows:
\begin{description}

\item[Symmetry:]
The rule set defining the lane changing of vehicles
can be both symmetric and asymmetric. The symmetric model is
interesting for theoretical considerations whereas the the asymmetric
model is more realistic.

\item[Stochasticity:]
The single lane model proved that a strictly deterministic
model is not realistic: the model did not show the desired spontaneous
formation of jams. In the case of
the two lane model the lack of stochasticity in combination with the
parallel update results in strange behaviour of slow platoons occupying
either lane: since none of the vehicles has reached its maximum
velocity and all evaluate the other lane to be better there is collective
change sidewise which is usually reversed over and over again until
the platoon dissolves or the platoon is passed by other vehicles.

We introduce stochasticity into the two lane rule set to reduce the
effective number of lane changes and thus dissolve those platoons.
The simulation also revealed that is effect is also
important in the asymmetric free--flow case (see \ref{pingpong}).

\item[Direction of Causality:]
In the single lane model a vehicle only looks ahead (= downstream = in
the direction of vehicle flow) so that causality can only travel
upstream (= in the opposite direction of
vehicle flow). A reasonable lane changing rule must include a check of
sites {\em upstream} in order not to disturb the traffic of the
destination lane. This would result in causality travelling downstream.

\end{description}

\noindent
A somewhat generic starting point for modeling passing rules is the
following: (T1) You look ahead if somebody is in your way.  (T2) You
look on the other lane if it is any better there.  (T3) You look back
on the other lane if you would get in somebody else's way.

Technically, we keep using $gap(i)$ for the number of empty sites
ahead in the same lane, and we add the definitions of $gap_o(i)$ for
the forward gap on the other lane, and $gap_{o,back}$ for the backward
gap on the other lane.  Note that if there is a vehicle on a
neighbouring site both return -1.  The generic multi-lane model then
reads as follows.
\begin{samepage}
A vehicle $i$ changes to the other lane if all of the following conditions
are fulfilled:
\begin{itemize}
\item $gap(i) < l$ ({\bf T1}),
\item $gap_o(i) > l_o$ ({\bf T2}),
\item $gap_{o,back}(i) > l_{o,back}$ ({\bf T3}), and
\item $rand() < p_{change}$ ({\bf T4}).
\end{itemize}
$l$, $l_o$, and $l_{o,back}$ are the parameters which decide how far
you look ahead on your lane, ahead on the other lane, or back on the
other lane, respectively.
\end{samepage}
\noindent
According to the before mentioned characteristics we associate
the parameters of our rule set:
\begin{center}
\begin{tabular}{c|cc}
\hline
characteristic & yes & no \\
\hline
symmetry     & {\bf T1} for L$\rightarrow$R & no {\bf T1} for
L$\rightarrow$R \\
stochasticity  & $prob_{c}<1$ & $prob_{c}=1$ \\
backward causality & $l_{o,back} > 0$ & $l_{o,back} = 0$ \\
\hline
\end{tabular}
\end{center}
\section{A lane changing model}

As an example, we start with $l=v+1$, $l_o = l$, $l_{o,back} = 5 =
v_{max}$, and $p_{change}=1$.  That means that both $l$ and $l_o$ are
essentially proportional to the velocity, whereas looking back is not:
It depends mostly on the expected velocity of other cars, not on one's
own.

In the symmetric version of this model, cars remain on their lane as
long as they don't ``see'' anybody else.  If they see somebody ahead
on their own lane (i.e.\ $gap < v+1$), then they check on the other
lane if they can switch lanes and do so if possible.  Afterwards, if
they are satisfied, they remain on this lane until they become
dissatisfied again.

In the asymmetric version, cars always try to return to the right
lane, independent of their situation on the left lane.

Space-time-plots both of the symmetric and the asymmetric version are
shown in Figs.~\ref{spacetime_symm_lb5} and~\ref{spacetime_asym_lb5}.
For these plots, we simulated a system with a length of 12,000 sites
of which we plot 400 sites in 400 consecutive time-steps. The density
is 0.09 which is slightly above the density of
maximum flow (see below). Vehicles go from left to right (spatial
axis) and from top to bottom (time axis). Traffic jams appear as solid
areas with steep positive inclination whereas free flow areas are
light and have a more shallow negative inclination. Each plot is split
into two parts: The left part containing the left lane and the right
part containing the right lane, respectively.

Note that plot \ref{spacetime_asym_lb5} (left lane) gives a good
impression of the great number of lane changes through the high
frequency of short vehicle life lines appearing and disappearing: These
are vehicles that temporarily leave the right lane to avoid an
obstacle. They go back to their old lane as soon as the obstacle has
been passed.  It will be confirmed quantitatively that indeed the
rate of lane changes is much higher for the asymmetric model than for
the symmetric model.

\subsection{Simulation Setup}

Before going on, we would like to describe our standard simulations
set-up for the following observations.  Note that quantitative
simulation results were obtained with a much larger system than the
qualitative space-time plots.  We simulated a system of
length
\begin{displaymath}
L~=~133,333~sites~\approx~1000~km
\end{displaymath}
with closed boundary conditions, i.e.\ traffic was running in a loop.
We started with random initial conditions, i.e.\ $N$\ cars were randomly
distributed
on both lanes around the complete loop with initial velocity $v_0=0$.

Since the system is closed, the average density per lane is now fixed at
\[
\langle \rho \rangle_L = {N \over 2 \, L} \ ,
\]
where the ``2'' stands for the number of lanes.

The simulation was then started, 1000 time steps were executed to let
the transients die out, and then the data extraction was started.  The
flow which is found in the fundamental diagrams is both space and
time\footnote{We gather statistics every fifth time-step only, since
subsequent time steps are correlated.} averaged, i.e.
\[
\langle j \rangle_{L,T} = {1 \over T} \, {1 \over L} \,
\sum_i^L \, \sum_t^{T/5} \, v(i,5t) \ .
\]
Values for lane change frequency and ping pong lane change frequency
(see below) are obtained by the same averaging procedure except that
statistics are gathered every time step, since {\em by definition}
ping pong lane changes occur in subsequent time steps.

We usually used $T = 5000$, and the same procedure was
repeated for each density found in the plots. With a resolution of
$\Delta\varrho=0.01$ an average plot took about 22 hours of
computation time on a Sparc 10 workstation.

\subsection{Flow behavior}

By comparing these models with each other and with earlier results, we
make the following observations (Fig.~\ref{fig_flow_lb5_c1_0} unless
otherwise noted):

(i) Both for the symmetric and the asymmetric version, maximum flow is
higher than twice the maximum flow of the single lane model
(Fig.~\ref{fig_single_multi}).
Which means that, in spite of the additional disturbances which the lane
changing behavior introduces into the traffic flow, the general
effects are beneficial, probably by diminishing large deviations from
``good'' flow patterns.

(ii) Both for the symmetric and the asymmetric version, the combined
2-lane flow reaches a maximum at $\rho_{jmax} \approx 0.08$, which is
at or near a sharp bend of the flow curve.

(iii) For the asymmetric model, flow on the left lane keeps increasing
slightly for $\rho > \rho_{jmax}$, but this is over-compensated by the
decreasing flow on the right lane.

(ii) and (iii) together lead one to the speculation that maximum flow
in the asymmetric case here actually is connected to a ``critical''
flow on the right lane and a ``sub-critical'' flow on the left lane.
Any addition of density beyond here leads to occasional break-downs on
the right lane and thus to a much lower flow there.  Obviously, such
interpretations would have to be clarified by further investigations,
and the word ``critical'' would have to be used with more care, such
as is pointed out in~\cite{ NaPa:1} for the single lane case.

(iv) For both lanes combined, the curves for symmetric and asymmetric
traffic actually look fairly similar. If the above interpretation is
right, this means that the overall density of maximum flow is a fairly
robust quantity, but one can stabilize one lane at a much higher
density if this density is taken from the other lane.

(v) At very low densities in the asymmetric case, flow on the left
lane, $j_{left}$, only slowly builds up.  This is to be expected,
since at least two cars have to be close to each other to force one of
them on the left lane, leading to a mean field solution of
$j_{left}(\rho) \sim \rho^2$ for $\rho \approx 0$.

(vi) For $\rho > 0.4$, flows on both lanes in the asymmetric models
are fairly similar and similar to the lane flows in the symmetric
models.

\subsection{Lane changing behavior}

To get some further insight into the lane changing dynamics,
Fig.~\ref{fig_change_lb5} shows the frequency of lane changing
both for the asymmetric and the symmetric model.

(i) Note that in the asymmetric case there is a sharp bend in the
curve, which is not found for the symmetric case.  This bend is also
near $\rho_{jmax}$, giving further indication that the dynamics above
and below $\rho_{jmax}$ are different.

(ii) For the symmetric case, lane changing occurs with less than half the
frequency compared to the asymmetric case.

(iii) In the symmetric case, the lane changing frequency per site for
small densities increases approximately quadratically up to rather
high densities, whereas the same quantity for the asymmetric model
grows approximately linearly already for fairly low densities.  
suggests that for the symmetric case a mean field description of
interaction, $P(change) \propto \rho^2$, would be valid up to
comparably high densities.  
obvious that this does not work.  Since the vehicles have a strong
tendency to be on the right lane, already a density of 0.04 per lane
would be a density of 0.08 if everybody were on the right lane.  Yet,
$\rho=0.08$ is known to be already a density of high interaction in
single lane traffic.  Since this high interaction tends to spread
vehicles out~\cite{ SchaSchr:1, Schreck.etc}, each additional vehicle
simply adds its own share of lane changes, making the relation roughly
linear.

(iv) The maximum number of lane changes occurs at densities much
higher than $\rho_{jmax}$. The lane changing probability {\em per
vehicle}, however, reaches a maximum below the critical density
(Fig.~\ref{fig_lanechangespercar}).

\subsection{Ping pong lane changes}
\label{pingpong}

An artifact of the so far described algorithm is easily recognizable
when one starts with all cars on the same lane, say the right one.
Assuming fairly high density, then all cars see somebody in front of
them, but nobody on the left lane.  In consequence, everybody decides
to change to the left lane, so that {\em all\/} cars end up on the
left lane.  Here, they now all decide to change to the right lane
again, etc., such that these coordinated lane changes go on for a long
time (``cooperative ping-pong effect'').  This effect has already been
observed by Nagatani for the much simpler two-lane model~\cite{
nagatani:det1, nagatani:det2}.

One way around this is to randomize the lane changing
decision~\cite{ nagatani:pchange}.  The decision rules remain the same
as above, but even if rules T1 to T3 lead to a yes, it is only
accepted with probability $p_{change}$.  With this fourth lane
changing rule, patterns like the above are quickly destroyed.

In order to quantify the effects of a different $p_{change}$,
simulations with $p_{change}=0.5$ were run.  The observations can be
summarized as follows:

(i) The flow-density curves are only marginally changed
(Fig.~\ref{fig_flow_lb5}).

(ii) The frequency of lane changes is decreased in general, but,
except for $\rho < \rho_{jmax}$ in the asymmetric case, by much less
than the factor of two which one would naively expect
(Fig.~\ref{fig_change_lb5}).  That means that usually there is a {\em
dynamic\/} reason for the lane change, that is, if it is not done in
one time step due $p_{change}<1$, then it is re-tried in the following
time step, etc.

(iii) To better quantify in how far a $p_{change} < 1$
actually changes the pattern of vehicles changing lanes back and forth
in consecutive time steps, we also determined the frequency of ``ping
pong lane changes'', where a car makes two lane changes in two
consecutive iterations.  Obviously, there are left-right-left (lrl) and
right-left-right (rlr) ping pong lane changes.

Fig.~\ref{fig_pingpong_lb5} shows that reducing the probability to change
lanes, $p_{change}$, from 1 to 1/2 has indeed a beneficial
effect: The number of ping pong lane changes decreases by about a
factor of five.

Yet, for the symmetric case, the frequency of ping pong lane changes
is more than an order of magnitude lower in both cases anyway.  This
indicates that in simulations starting from random initial conditions,
the cooperative effect as described further above as cooperative
ping-pong effect does not really play a role for the statistical
frequency, because this effect should be the same for the symmetric
and the asymmetric model.  Instead, the cause of the ping pong lane
changes in the asymmetric model is as follows: Assume just two cars on
the road, with a gap of 5 between them.  With respect to velocity,
both cars are in the free driving regime, and their velocities will
fluctuate between 4 and 5.  Now assume that the following car has
velocity 5 from the last movement.  That means that it looks 6 sites
ahead, sees the other car, and changes to the left lane.  Then, assume
that in the velocity update, the leading car obtains velocity 5 and
the following car obtains velocity 4.  Then, after the movement step,
there is now a gap of 6 between both cars, and in the lane changing
step, the follower changes back to the right lane.  And this can
happen over and over again in the asymmetric model, but will not
happen in the symmetric model: Once the following car in the above
situation has changed to the left lane, it will remain there until it
runs into another car on the left lane.

To investigate this second kind of ping pong lane changes we ran
simulations recording whether a ping pong lane change was made at low
velocities $0 \le v \le 3$ or high velocities $4 \le v \le
5$. Fig.~\ref{fig_slowfastpingpong} shows a very distinct peak for
{\em fast} ping--pong--changes at low densities whereas {\em slow}
ping--pong--changes have a lower peak at higher densities similar to
that of the symmetric case.

This gives a strong indication that most lane changes are actually
caused by the ``tailgating effect'' as described above, which is an
artifact of the rules.  It is, though, to be expected that this
behavior does not have a strong influence on the overall dynamics: It
mostly happens in the free driving regime; as soon as, for example,
another car is nearby in the left lane, it is suppressed by the
looking back and forward on the other lane.

\subsection{Other Parameter Combinations}

We would like to mention two other parameter combinations. They are
presented because they generate artifacts which contradict the common
sense one would apply to the phenomena of traffic flow.

(i) In the first case the lookahead is reduced to $l=v$ instead of
$l=v+1$. While this change is negligible for vehicles at higher
velocities it becomes crucial to a vehicles stopped in a jam: assuming
the current velocity to be zero the vehicle looks {\em zero} sites
ahead and decides to remain in the current lane due to the
non--fulfilled rule $T1$. This state will persist until the
predecessor moves even if the other lane is {\em completely} free!
Fig.~\ref{fig_model1} shows the impact the reduced look ahead on
overall flow: for density $\rho>0.75$ there is no perceptible flow in
the right lane which corresponds to a traffic jam that occupies more or
less the whole right lane.

(ii) In the second case the look-back is reduced to $l_{o,
back}=0$. Vehicles no longer check whether their lane changing could
have a disadvantageous effect on the other lane which corresponds to a
very egoistic driver behaviour. Fig.~\ref{fig_flow_lbx_c1_0} shows flow
density relationships for look-back $l_{o, back}=5$ and $l_{o,
back}=0$. It is obvious that the decrease in look-back also decreases
the maximum flow at critical densities. Moreover $l_{o,back}=0$
seems to split the curves of the symmetric and asymmetric
cases which used to be almost identical for $l_{o,back}=5$: the lack
of look-back is much more disadvantageous for asymmetric than for
symmetric rules.

In Figs.~\ref{spacetime_symm_lb0} and~\ref{spacetime_asym_lb0} we used
$l_{o,back}=0$ for the symmetric and asymmetric rule sets with one plot
per lane.  It is clearly visible (compare to
Figs.~\ref{spacetime_symm_lb5} and~\ref{spacetime_asym_lb5}) how
$l_{o,back}=0$ completely disrupts the laminar flow regime.  Vehicles
change lanes without looking back; and due to the formulation of the
model this does not cause accidents, but causes the obstructed
vehicles to make sudden stops.  Since these stops are caused more or
less randomly, the regime becomes much more randomly disturbed than
before, somewhat reminiscent of the Asymmetric Stochastic Exclusion
Process (see~\cite{ Schreck.etc, Nagel.95.flow}).

As seen before the effect is even more drastic for the asymmetric
rule set since the number of lane changes is higher than in the
symmetric case. In Fig.~\ref{spacetime_asym_lb0} with $l_{o,back}=0$
dynamics are dominated by small traffic jams caused by lane changes,
while in Fig.~\ref{spacetime_symm_lb5} there are still some fairly
laminar areas.

\section{Discussion}

Compared to reality (e.g.~\cite{  McDonald.Brackstone}), the lane change
frequency in the asymmetric models presented here is by about a factor
of 10 too high.  Using $p_{change}=0.1$ would correct this number, but
is dynamically not a good fix: It would mean that a driver follows a
slower car in the average for 10~seconds before he
decides to change lanes.  Besides, it was shown that about 90\% of the
lane changes in the asymmetric models here are produced by an
artificial ``tailgating dance'', where a follower changes lanes back
and forth when following another car.  It remains an open question in
how far artifacts like this can be corrected by the current modeling
approach or if it will be necessary to, e.g., introduce memory: If
one remembers to just have changed lane from right to left, one will
probably stay on that lane for some time before changing back.

Another defect of the models presented in this paper is that the
maximum flow regime is most probably represented incorrectly.  Both
measurements (e.g.~\cite{ McDonald.Brackstone} or Fig.~3.6
in~\cite{ May}) or everyday observation show that real traffic shows a
``density inversion'' long before maximum flow, that is, more cars
drive on the left lanes than on the right lanes.  This effect is more
pronounced for countries with higher speed limits.  Let us denote by
$\rho^{1lane}_{jmax}$ the density of maximum flow of the {\em single
lane\/} case.  It follows for the real world two lane case that at a
certain point the left lane will have a density higher than this
density $\rho^{1lane}_{jmax}$ whereas the right lane has a density
lower than $\rho^{1lane}_{jmax}$.  When further increasing the overall
density, then the flow on the left lane will decrease whereas it still
increases on the right lane.  It is unclear if the net flow here
increases or decreases; but it should become clear that instabilities
here are caused by the left lane first.  This is in contrast to the
models of this paper, where the right lane reaches the critical
density first.  Work dealing with this problem is currently in
progress.

Also the effect of different maximum velocities will be addressed in
later papers.

\section{Summary and conclusion}

We have presented straightforward extensions of the cellular automata
approach to traffic flow so that it includes two-lane traffic. The
basic scheme introduced here is fairly general, essentially consisting
of two rules: Look ahead in your own lane for obstructions, and look
in the other lane if there is enough space.  The flow-density
relations of several realizations of this scheme have been
investigated in detail; possible artifacts for certain parameter
choices have been pointed out.  In general, there seem to be two
important lessons to be drawn from our investigations:

\begin{itemize}
\item
Checking for enough space on the other lane (``look-back'') is
important if one wants to maintain the dynamics consisting of laminar
traffic plus start--stop waves which is so typical for traffic.

\item
Especially in countries with high speed limits, observations show a
density inversion near maximum flow, that is, the density is higher on
the left lane than on the right lane.  This effect is not reproduced
by our models (work in progress).

\end{itemize}
Yet, in general, it seems that the approach to multi lane traffic
using simple discrete models is a useful one for understanding
fundamental relations between microscopic rules and macroscopic
measurements.

\section{Acknowledgements}
We thank A.~Bachem and C.~Barrett for supporting MR's and KN's work as
part of the traffic simulation efforts in Cologne (``Verkehrsverbund
NRW'') and Los Alamos (TRANSIMS).  We also thank them, plus
Ch.~Gawron, T.~Pfenning, S.~Rasmussen, and P.~Wagner for help and
discussions.  Computing time on the Parsytec GCel-3 of the Zentrum
f\"ur Paralleles Rechnen K\"oln, on the Paragon XP/S 5 and XP/S 10 of
the Zentralinstitut f\"ur Angewandte Mathematik (ZAM) of the
Forschungszentrum J\"ulich and on the SGI-1 of the Regionales
Rechenzentrum K\"oln is gratefully acknowledged. We further thank all
persons in charge of maintaining the above mentioned machines.

The work of MR and KN was supported in part by the ``Graduiertenkolleg
Scientific Computing K\"oln/St.~Augustin''.


\def\jrn{\em}
\def\vol{\bf}
\def\book{\em}

\def\ESM{in {\proc Proc.\ 1993 European Simulation
Multiconference}, edited by A.~Pave (1993)}

\newpage


\emptyeps{timespaceplot_800x400_symm_lb5_d0.09.ps}{symmetric,
$l_{o,back}=5$, left + right lanes}{spacetime_symm_lb5}

\emptyeps{timespaceplot_800x400_asym_lb5_d0.09.ps}{asymmetric,
$l_{o,back}=5$, left + right lanes}{spacetime_asym_lb5}

\eps{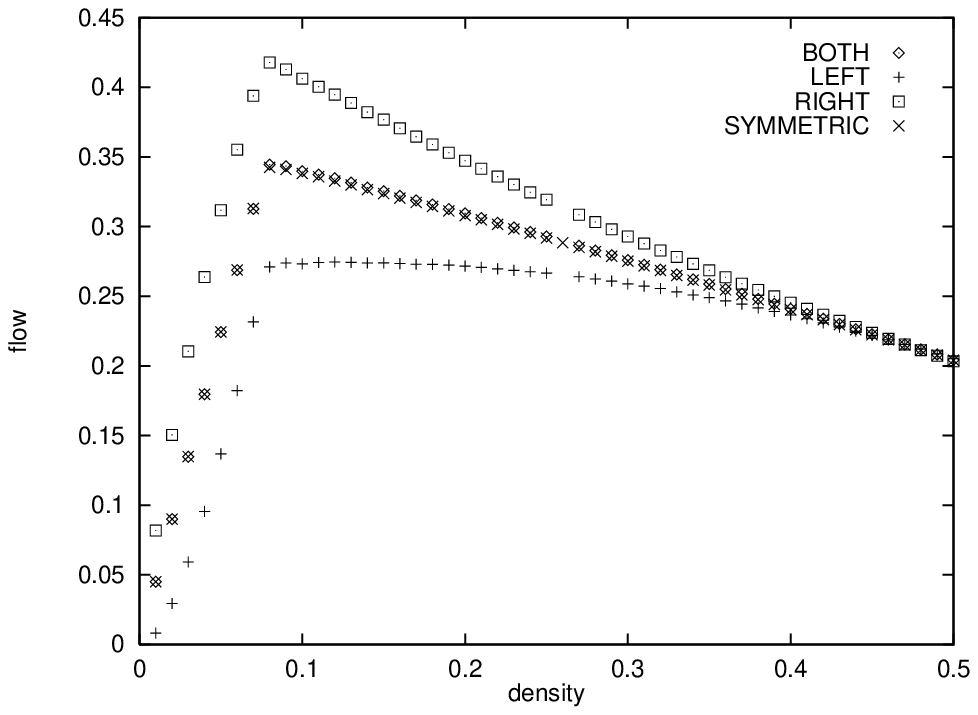}{Flow $LookBack=5$,
$p_{change}=1.0$}{fig_flow_lb5_c1_0}

\eps{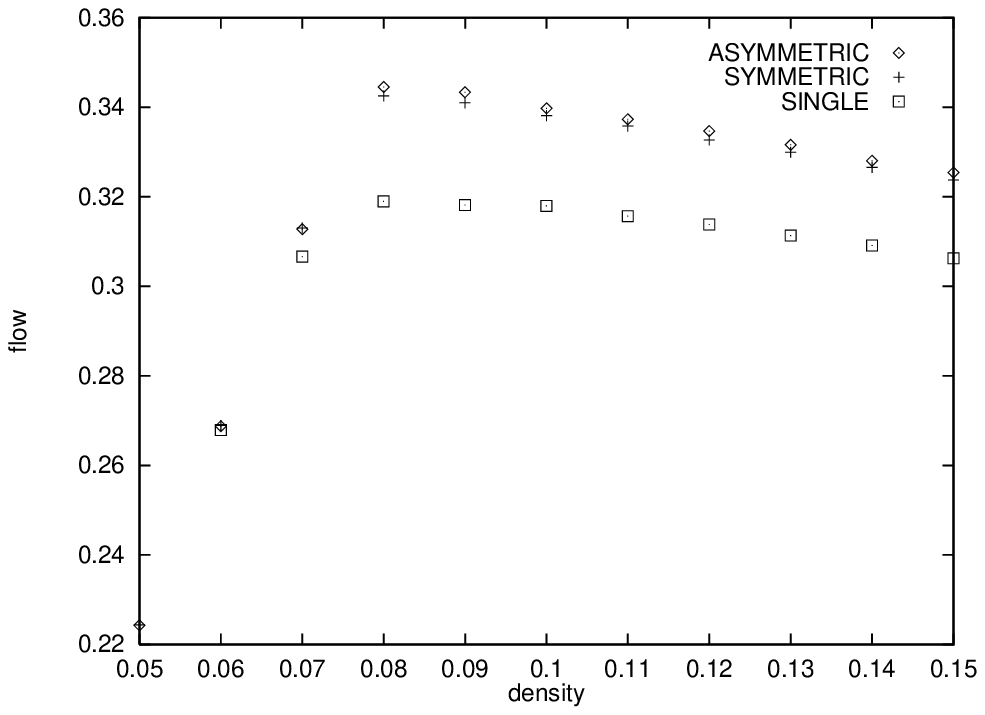}{Comparison Single Lane to Multi
Lane}{fig_single_multi}

\eps{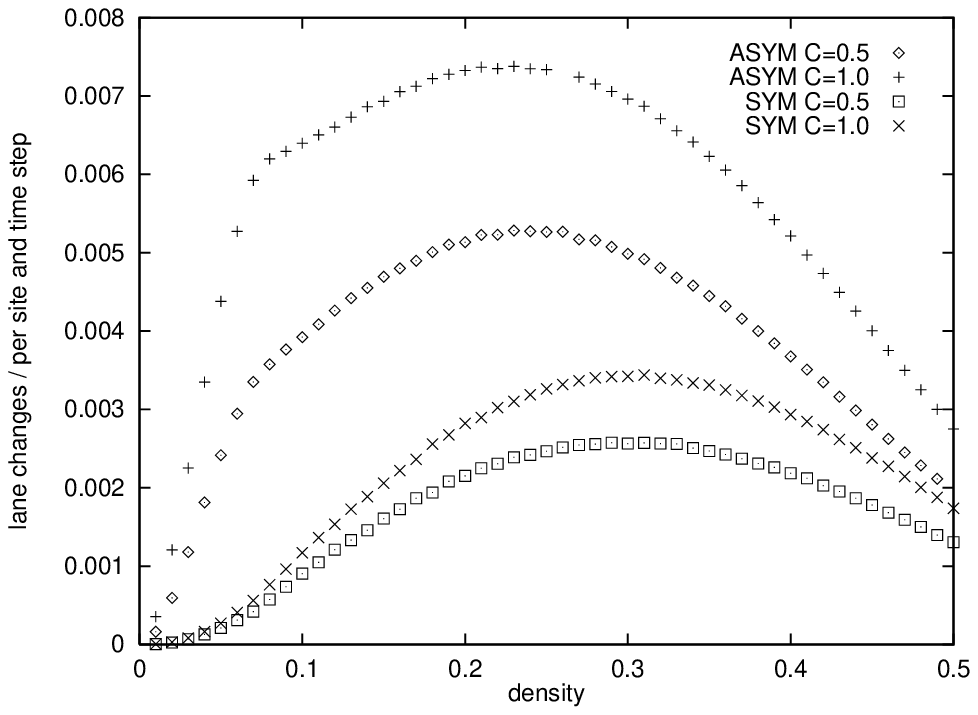}{Simple Lane Changes $LookBack=5$}{fig_change_lb5}

\eps{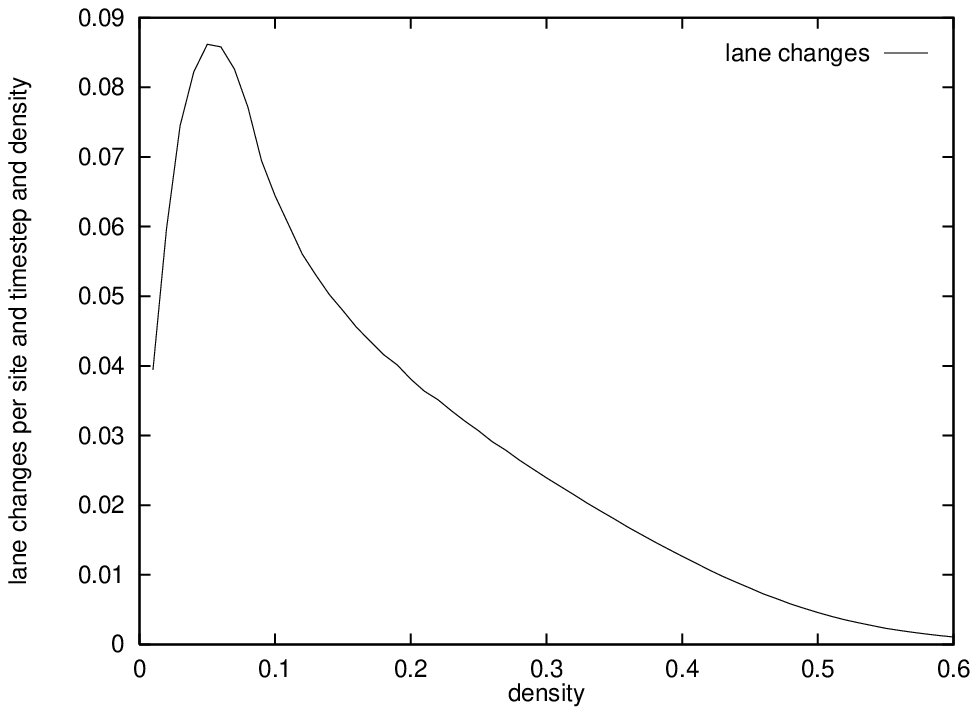}{Lane Changes Per Vehicle $l=5$,
$p_{change}=1.0$}{fig_lanechangespercar}

\eps{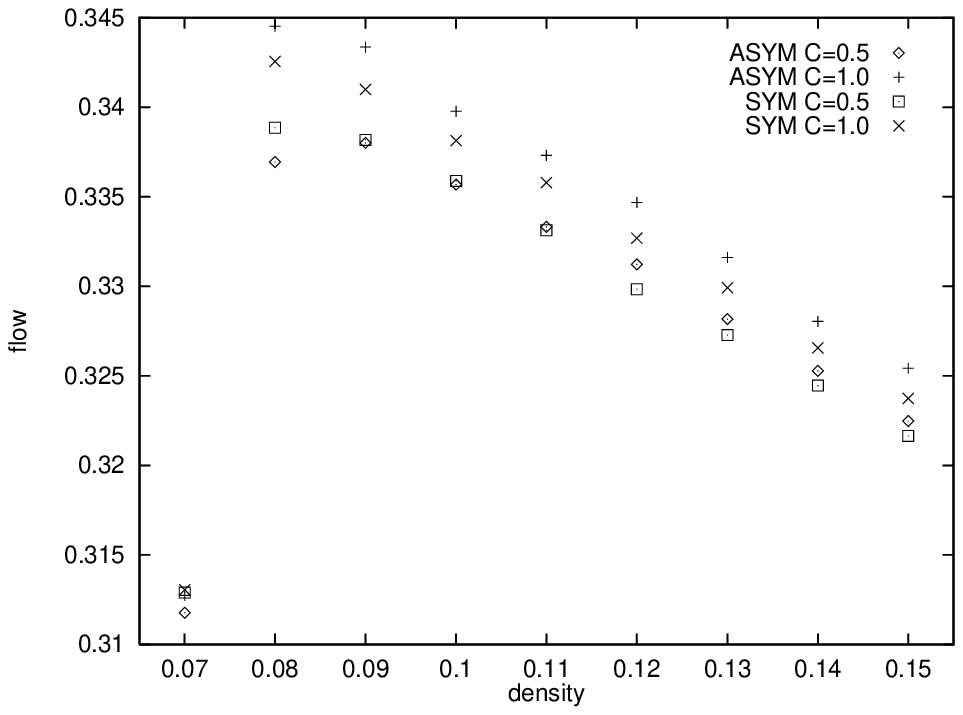}{Flow $LookBack=5$, different $p_{change}$}{fig_flow_lb5}

\eps{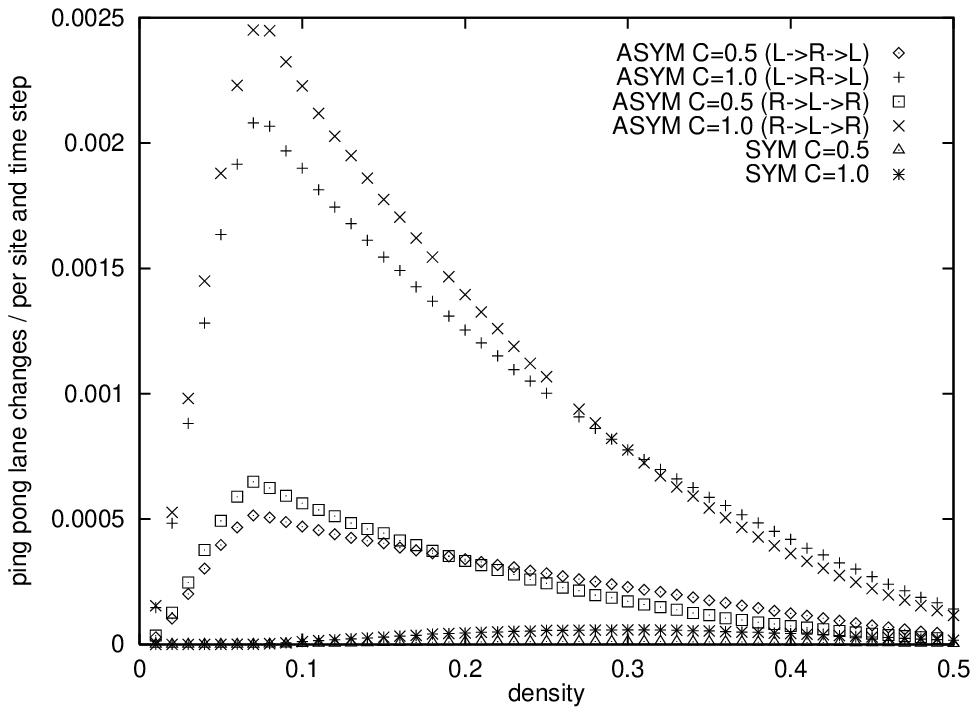}{Ping Pong Lane Changes $LookBack=5$,
different $p_{change}$}{fig_pingpong_lb5}

\eps{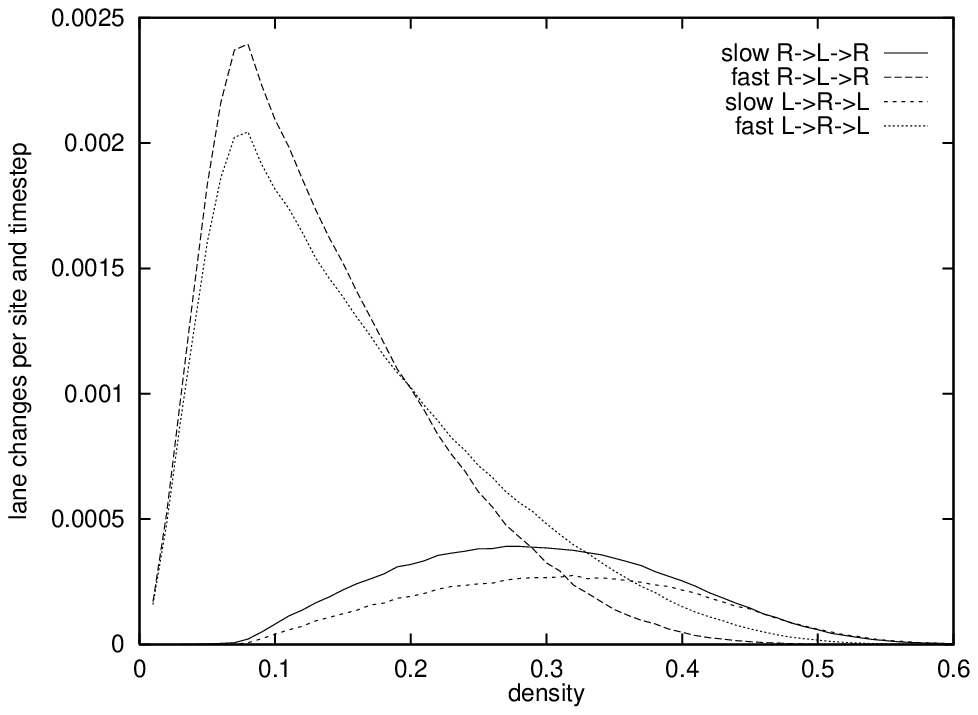}{Slow and Fast Ping Pong Lane
Changes}{fig_slowfastpingpong}

\eps{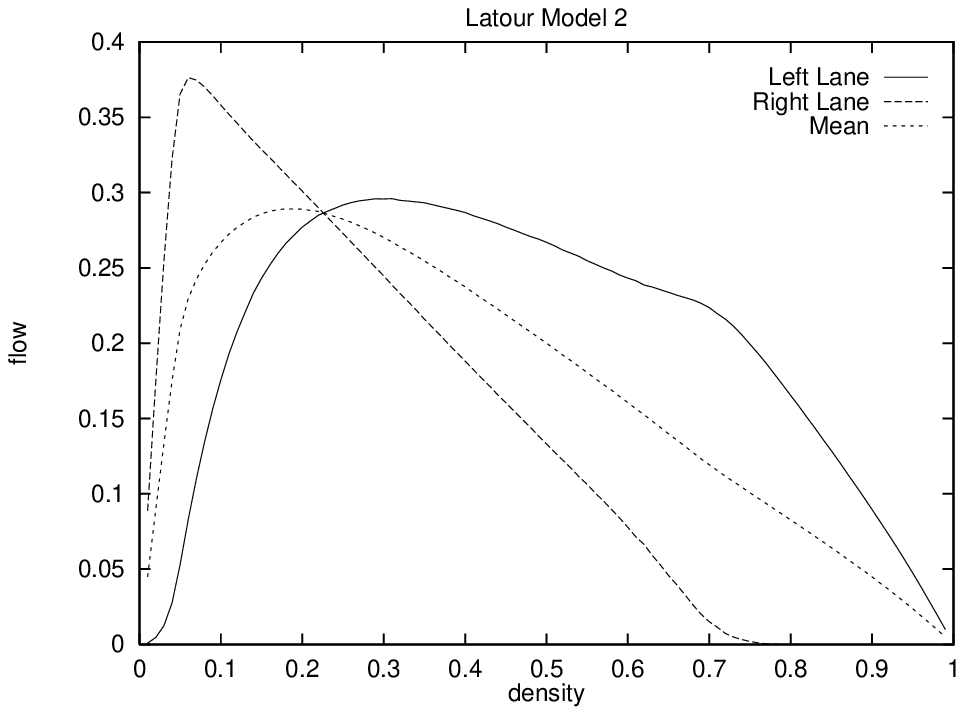}{Flow with $l=v$}{fig_model1}

\eps{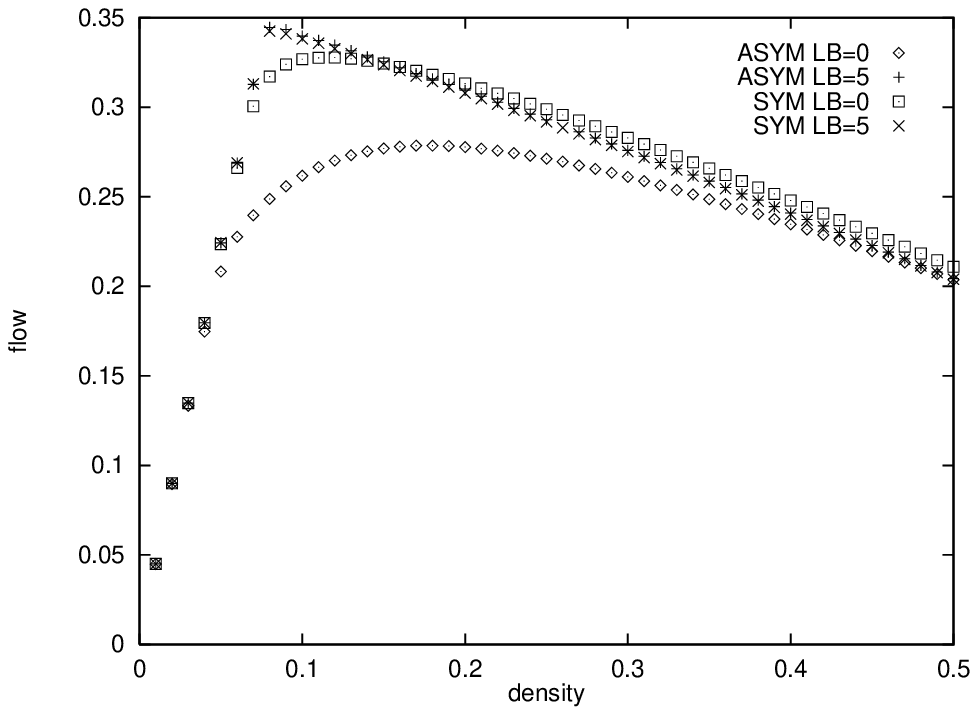}{Flow $p_{change}=1.0$, different
$LookBack$}{fig_flow_lbx_c1_0}

\emptyeps{timespaceplot_800x400_symm_lb0_d0.09.ps}{symmetric,
$l_{o,back}=0$, left + right lanes}{spacetime_symm_lb0}

\emptyeps{timespaceplot_800x400_asym_lb0_d0.09.ps}{asymmetric,
$l_{o,back}=0$, left + right lanes}{spacetime_asym_lb0}
\end{document}